\documentstyle[aaspp4,flushrt,psfig]{article}

\begin{document}
\title{The Blackhole-Dark Matter Halo Connection}

\author{Bassem M. Sabra$^1$, Charbel Saliba$^{2}$, 
Maya Abi Akl$^{2}$, and Gilbert Chahine$^{2}$}

\affil{$^1$Department of Physics \& Astronomy, Notre Dame University-Louaize,
POB 72 Zouk Mikael, Zouk Mosbeh, LEBANON; email: bsabra@ndu.edu.lb
\\
$^2$Department of Physics, Lebanese University II, Fanar, LEBANON
}
\begin{abstract}
We explore the connection between the central
supermassive blackholes (SMBH) in galaxies and the dark matter halo
through the relation between the masses of the SMBHs and the maximum
circular velocities of the host galaxies, as well as the relationship
between stellar velocity dispersion of the spheroidal component and the
circular velocity. Our assumption here is that the circular velocity
is a proxy for the mass of the dark matter halo. We rely on a
heterogeneous sample containing galaxies of all types. The only
requirement is that the galaxy has a direct measurement of the mass of
its SMBH and a direct measurement of its circular velocity and its
velocity dispersion. Previous studies have analyzed the connection
between the SMBH and dark matter halo through the relationship between
the circular velocity and the bulge velocity dispersion, with the
assumption that the bulge velocity dispersion stands in for the mass of
the SMBH, via the well{}-established SMBH mass{}-bulge velocity
dispersion relation. Using intermediate relations may be
misleading when one is studying them to decipher the active ingredients
of galaxy formation and evolution. We believe that our approach will provide 
a more direct probe of the SMBH and the dark matter halo connection. We 
find that the correlation between the mass of supermassive blackholes and 
the circular velocities of the host galaxies is extremely weak, leading us to 
state the dark matter halo may not play a major role in regulating the 
blackhole growth in the present Universe. 

\end{abstract}

\section{Introduction}

Data has been accumulating over the last several years on an increasing
number of galaxies in terms of measurements of the bulge velocity
dispersion $\sigma$, mass of the central supermassive blackhole
(SMBH) $M_{bh}$, and the circular velocity, $v_{c}$, of the host galaxy 
(McConnell \& Ma 2013; Hu 2009; Graham 2008; Hu 2008; Ho 2007; 
Courteau et al. 2007; Pizella et al. 2005; 
Baes et al. 2003; Ferrarese 2002; Tremaine et al. 2002; Onken et al 2004; 
Peterson et al. 2004; Nelson et al. 2004; Merrit and Ferrarese 2001; Bedregal et
al. 2006; Bernardi et al. 2002; Kronawitter et al. 2000; Palunas et al. 2000; 
Prugniel et al 2001; Verheijen et al. 2001). With the increase in the number of the 
galaxies for which $\sigma$, $v_c$, and $M_{bh}$ have been measured, 
it is becoming possible to ask, and attempt to answer, questions about 
the processes that govern the formation of galaxies and their central SMBHs, 
and the symbiotic relationship between an SMBH and its host galaxy. One 
way to approach these issues is to compare observed correlations, or lack 
thereof, between $v_{c}$, $\sigma$, and $M_{bh}$ with theoretical predictions 
obtained from models of galaxy/SMBH formation (e.g., Di Matteo et al. 2008; 
Di Matteo et al. 2003; Haehnelt \& Kauffmann 2000; Kauffmann \& Haehnelt 
2000). 

The observed $M_{bh}-\sigma$ relation is now on a firm basis
(Beifiori et al. 2012; Gultenkin et al. 2009; Hu 2009; Graham 2008; Hu 2008; 
Tremaine et al. 2002; Ferrarese \& Merritt 2000; Gebhardt et al. 2000) 
and it points to a common history between the 
SMBH and the spheroidal component of galaxies. Moreover, it was 
recently found that $v_{c}$ and $\sigma$ are correlated 
(Courteau et al. 2007; Baes et al. 2003, Ferrarese 2002). However, 
Ho (2007) by a using a sample of 792 galaxies have shown 
that the $v_c-\sigma$ relation has a very large scatter, and it depends 
on galaxy morphology. 

Ferrares (2002) investigated the possibility of the masses of the supermassive 
blackhole being correlated with the total gravitational mass of the host galaxy or 
the mass of the dark matter halo. The author found a tight correlation between 
$v_c$ and $\sigma$ for a sample of 36 galaxy, and by using the well-known 
$M_{bh} - \sigma$ relation, and connecting the mass of the dark matter 
halo with the circular velocity of the galaxy, report a correlation between 
the mass of the supermassive blackhole and the circular velocity of 
the host galaxy. Baes et al. (2003) reported a relation between $M_{bh}$ 
and $v_{c}$. The rationale behind this correlation is that both quantities 
depend on the mass of the dark matter halo. A heavier halo will result in a 
deeper potential well that will give rise to a higher $M_{bh}$, and at the same time 
lead to a higher $v_{c}$. Zasov et al. (2005), on the other hand, found a very weak 
correlation between $M_{bh}$ and $v_c$. However, the $M_{bh}$ values 
in both Baez et al. (2003) and Zasov et al. (2005), as well as in Ferrarese (2002), 
were obtained from the $M_{bh}-\sigma$ relation. 
Ho et al. (2008) report a correlation between $M_{bh}$ and $v_c$ for a 
sample of active galaxies, where the SMBH masses were measured using 
the virial method (Kaspi et al. 2000; Greene \& Ho 2005).  
More recently, Kormendy et al. (2011) used dynamical $M_{bh}$ measurements 
to find that $M_{bh}-v_{c}$ relation is very weak in a sample of 25 spiral galaxies.
The same result is confirmed in Beifiori et al. (2012) who used  a large sample, 
105 galaxies, but the used $M_{bh}$'s  are a mix of dynamical 
measurements from Gultekin et al. (2009a) and upper limits from 
Beifiori et al. (2009), with the upper limits taken as surrogate of black mass. 
{\bf More recently, Sun et al. (2013) analyzed the $M_{bh}-v_{c}$ relation for 
22 galaxies with $v_{c}$ determined from \ion{H}{1} observations, and came up 
with a broad relation with large instrinsic scatter.}

Given the controversy surrounding the 
$M_{bh}-v_{c}$ relation and its strength, using relations to find blackholes 
masses will only introduce additional scatter and makes it more 
difficult to reach firm conclusions for or against a genuine relation. 
Most of the previous studies that have investigated the relation 
between $M_{bh}$ and the circular velocity did so indirectly since 
they "measured" the SMBH masses through the $M_{bh}-\sigma$ 
relation for galaxies with measured $v_{c}$ and $\sigma$. The danger 
in the indirect approach is that one is at 
the risk of masking, or leading to, a correlation between 
$M_{bh}-v_{c}$ through the $v_{c}-\sigma $ and $M_{bh}-\sigma$, 
and it necessarily ignores bulge{}-less spirals with SMBHs, further 
masking the true connection between the SMBH and the dark matter 
halo. 

The aim of this paper is to present a direct study of the nature of the 
$M_{bh}-v_c$ relation for {\bf 53} galaxies with dynamically measured $M_{bh}$. 
This is the largest sample to date of galaxies that have these two properties 
available. We also present, as a by-product of our study,  the $M_{bh}-\sigma$ 
relation based on {\bf 89} galaxies. We are revisiting the question of the 
$M_{bh}-v_c$ relation with a slightly "cleaner" sample of blackhole masses. 
It is our hope here that by using dynamical $M_{bh}$'s we will be able to 
probe directly the intrinsic relations between a SMBH and the dark matter halo 
of the host galaxy. 


\section{Description of the Sample}
Our main objective is to build a sample of galaxies that have dynamical 
$M_{bh}$ and $v_c$ measurements. In order to make sure that we are 
not missing any objects we collected a large sample of 376 galaxies of all types 
for which measurements of $M_{bh}$, $v_{c}$, and/or $\sigma$ exist. 
This large sample was culled from papers that present  measurements 
of SMBH masses (McConnell \& Ma 2013; Gultenkin et al. 2009a; Pastorini et al 2007; 
de Fransesco et al. 2006; Ferrarese \& Ford 2005; Onken et al. 2004; 
Peterson et al. 2004; Nelson et al. 2004; Gebhardt et al. 2000; Kaspi et al. 2000; 
Richstone 1998), papers that study the $M_{bh}-\sigma$ relation 
(Hu 2009; Gultekin et al. 2009a, Graham 2008; Hu 2008; Tremaine et al. 
2002; Gebhardt et al. 2000; Ferrarese \& Merritt 2000), and papers 
that deal with the $v_{c}-\sigma$ and/or $M_{bh}-v_{c}$ relationships 
(Kormendy et al. 2011; Courteau et al. 2007; Ho 2007;  
Pizella et al. 2005; Baes et al. 2003; Ferrarese 2002). 


We found 342 galaxies with $\sigma$ measurements, 269 galaxies with 
$v_{c}$, and 125 galaxies with $M_{bh}$. We reject 35 out of the 125 
blackhole masses since they are determined via reverberation mapping, 
which is caliberated to the $M_{bh}-\sigma$ relation (Onken et al. 2004; 
Peterson et al. 2004). The intersections between 
these subsets resulted in 53 galaxies that have dynamical $M_{bh}$ 
and $v_c$,  89 galaxies with both dynamical $M_{bh}$ and $\sigma$, and 
251 galaxies that have both $\sigma$ and $v_{c}$. 

When error bars on a quantity for a particular galaxy are not available
we assign an error by multiplying that quantity by the 
average percent error derived for that quantity from all 
galaxies that have errors. Out of 342 galaxies that have $\sigma$ values, 
332 have errors listed, and 251 out of 269 galaxies have errors on $v_c$. 
The average percent error on $\sigma$ is $\sim 10$\% and on $v_{c}$  
is $\sim 5$\%. All available SMBH dynamical mass measurements have 
error bars quoted in the literature. The errors bars on $M_{bh}$ are not 
symmetric. We calculate the average of the interval and assign this 
error to the data point when peforming the fits. 

We list in Table 1 our final sample of 89 galaxies with dynamically-determined 
SMBH masses, stellar velocity dispersions, and circular velocities (53 out 
of 89). All of the $\sigma$ values are from HyperLeda$^1$ 
(Paturel et a. 2003) except  for IC 2560 (Garcia-Rissman 
et al. 2005) and Cygnus A (Graham 2008), and MW (Baes et al. 2003; Ferrarese 2002). 
Most of the blackhole masses are from Hu (2009) and McConnell \& Ma (2013) except for 
NGC 1300; NGC 2748; NGC 2778; NGC 4342; NGC 4374;  NGC 4945; NGC 7582 (Graham 2008), 
NGC 4303 (Pastorini 2007),  NGC 4486 Kormendy et al. (1996), NGC 4594 (Kormendy et al. 1988). 
The distances in Table 1 are from the above-mentioned papers. We include in Table 1 references 
to the original papers and the methods used to measure the blackhole masses. 

We use circular velocities from 
optical rotation curves whenever available: NGC 2787,  
NGC 1023, NGC 3115, NGC 3384,  and NGC 4649 (Neistein et al. 1999);  
NGC 1399, NGC 3379, NGC 4374, NGC 4486, and 
NGC 4486B (Kronawitter et al. 2000), and NGC 5846 (Pizzella et al. 2005).  
We use circular velocities derived from \ion{H}{1} line widths as 
listed in Courteau et al. (2007) for: MW, NGC 224, and NGC 4258 
(Ferrarese et al. 2002); NGC 2974 (Pizzella et al. 2005); and 
NGC 3031, NGC 3227, NGC 4303, and NGC 4594 (Prugniel et al. 2001). 
We use at face value the circular velocities in HyperLeda for the remaining 
34 galaxies. They are also based on \ion{H}{1} line widths, which brings 
the number of galaxies in our sample with $v_c$ derived from \ion{H}{1} 
line widths to 42. We note here that the results that we arrive at below 
do not depend on the origin of the circular velocities.

\section{Analysis and Discussion}
Studying the correlations involves fitting straight lines in
log{}-log space: $log y= \alpha + \beta log x$.  
The results of the fitting have been known to depend in 
part on the details of the linear regression analysis (Tremaine et al.
2002). For the sake of definiteness, we limit
ourselves to the prescription of Press et al. (1992) for data
with symmetric error bars on both variables. 
{\bf The errors on $\alpha$ and $\beta$ are $1-\sigma$ fitting errors 
when $\chi^2$ differs by unity from its mininum value.} A shortcoming of this
approach is that it does not take into account the presence of
intrinsic dispersion in the data. We follow the quick fix offered by
Tremaine et al. (2002): we add in quadrature a parameter to the error of 
the y{}-coordinates (Gebhardt et al. 2000). This parameter is a measure 
of the intrinsic dispersion. Its value is adjusted by hand to lead a 
reduced chi{}-squared of unity. {\bf This prescription (Tremaine et al. 2002)
is essentially implemented in FITEXY and, its more advanced variant MPFITEXY 
(Markwardt 2009; Williams et al. 2010), which is what we use in this paper.}


\subsection{A $M_{bh}-v_{c}$ Relation?}
We plot in Fig. 1 $\log (M_{bh}/10^{8}M_{{\odot}})$ against 
$\log (v_{c}/200\ km\ s^{-1})$ for 53 galaxies, made up of 13 ellipticals, 
16 lenticulars, and 24 spirals. (Table 1). It is clear that 3 galaxies 
whose $v_c \lesssim 100$~km~s$^{-1}$ don't fall in the region where the 
other data points cluster. These 3 galaxies are IC 1459 (E3), NGC 5252 (S0), 
and NGC 3608 (E2). We do not include them in our subsequent analysis 
when we fit:
\begin{equation}
\log \frac{M_{bh}}{10^{8}M_{{\odot}}}=\alpha + \beta \log \frac{v_{c}}{200\ km\ s^{-1}} 
\end{equation}
to the data points in Fig. 1. 

A casual inspection of Fig. 1 shows that the correlation between the $M_{bh}$ 
and $v_c$, if there is any,  is very weak.  Taken separately, galaxies belonging 
to the same morphological type form essentially scatter plots and at best barely 
show a hint of a correlation. A visual inspection in Fig. 1 that does not distinguish 
between the galaxy types can be misleading. An apparent correlation will appear 
to be present if one is not careful. The spirals, represented as blue crosses, have 
on average lower blackhole masses and velocities than the ellipticals and the 
lenticulars. Placed on the same plot, this offset between spiral and E/S0 galaxies 
makes one to believe that there is a correlation. 

To be more quantitative, we use the procedure outlined above to obtain:

\begin{equation}
\log \frac{M_{bh}}{10^{8}M_{{\odot}}}=(-0.25\pm 0.11) + (2.28\pm 0.67) \log \frac{v_{c}}{200\ km\ s^{-1}}, 
\end{equation}
with an intrinsic scatter of  0.75 dex. The intrinsic scatter is very large and so 
is the error on the slope. This leads to conclude that there is no 
correlation between the mass of supermassive blackhole and the 
circular velocity of the host galaxy. {\bf This result is broadly consistent with those of 
Sun et al. (2013), Beifiori et al. (2012), and Kormendy et al. 2011}.


Previous authors have studied the relation between $M_{bh}$ and $v_c$ 
and their conclusions are varied. Baes et al. (2003) reported a 
slope of $4.21\pm 0.60$ for a sample of 40 galaxies. 
However, Baes et al. (2003) determined 
the masses of the blackholes indirectly by using the well-established 
$M_{bh}-\sigma$ relation of Tremaine et al. (2002). They combined 
their $v_c-\sigma$ relation with the $M_{bh}-\sigma$ relation of Tremaine 
et al. (2002) to derive a $M_{bh}-v_c$ relation. Ho (2007) using a sample of 
792 galaxies has shown that the $v_c-\sigma$ relation has appreciable 
variation depending on host galaxy properties. 
Ho (2007) cautions against replacing the bulge, 
i.e. stellar velocity $\sigma$, with the halo, $v_c$, in attempts to use 
$v_c$ to derive blackhole masses based on a $M_{bh}-v_c$ relation.
The slope arrived at by Baes et al. (2003) is in 
agreement with the predictions of cosmological simulations (Di Matteo 
et al. 2003), but it is an artefact of combining the $M_{bh}-\sigma$ 
relation with the less accurate $v_c-\sigma$ relation.  Zasov et al. (2005) 
found significant scatter in the $M_{bh}-v_c$ plots 
and concluded that the relation between $M_{bh}-v_c$ is very weak. 
Their sample of 41 galaxies suffered contamination from galaxies, 20 
in number, whose blackhole masses were determined indirectly from 
the $M_{bh}-\sigma$ relation. Sabra et al. (2008) reported a slope 
of $6.75\pm 0.8$ based on a small sample of only 16 galaxies. 

Bandara et al. (2009) reported a correlation between 
$M_{bh}$ and $M_{tot}$, the total gravitational mass of the host galaxy.  
The total mass was determined from gravitational lensing observations, 
and their blackhole masses were determined indirectly through the 
$M_{bh}-\sigma$ relation (Gultenkin al. 
2009a). However, when they calculate $M_{bh}$ through the $M_{bh}-n$ 
relation (Graham et al. 2001; 2007), where $n$ is the Sersic index, 
they find no signficant correlation between $M_{bh}$ and $M_{tot}$. 

We derive here the expected $M_{bh}-v_c$ relation from the Bandara et al. 
(2009) $M_{bh}-M_{tot}$ relation: 
\begin{equation}
\log M_{bh}=(8.18\pm 0.11)+(1.55\pm 0.31)(\log M_{tot}-13). 
\end{equation}
We use eq. (5) in Ferrarse (2002), in which the author assumes that $v_c$ is 
equal to the virial velocity and uses relations from Bullock et al. (2001) between the 
mass of the dark matter halo and the circular velocity, We further assume 
that the total gravitational mass is equal to the mass of the dark matter halo to get: 
\begin{equation}
\log \frac{M_{DM}}{10^{12} M_\odot}=0.15+3.32 \log \frac{v_c}{200 km s^{-1}}, 
\end{equation}
Combining eqs. (3) and (4) above we obtain the following:
\begin{equation}
\log \frac{M_{bh}}{10^8 M_\odot}=-1.14+5.15 \log \frac {v_c}{200 km s^{-1}}. 
\end{equation} 
Eq. (5) is semi-observational in the sense that it depends on an observational 
relation eq. (3) that connects the mass of the blakchole to the total gravitational 
mass of the galaxy, and on the theoretical relation eq. (4) that connects the 
mass of the dark matter halo to the circular velocity of the galaxy. 

We also derive a theoretical counter-part to Eq. (5) by using
\begin{equation}
\log \frac{M_{bh}}{10^8 M_\odot}=-0.15+1.33 \log \frac{M_{DM}}{10^{12} M_\odot}, 
\end{equation}
reported in di Matteo et al. (2003) from cosmological simulations, combined 
with the theoretical prescription of Ferrarese (2002) derived from 
Bullock et al. (2001), eq. (4) above, to obtain:
\begin{equation}
\log \frac{M_{bh}}{10^8 M_\odot}=0.04+4.43 \log \frac {v_c}{200 km s^{-1}}. 
\end{equation} 

We overplot eqs. (5) and (7) in Fig. 1. Most of the 46 galaxies fall between these two 
relations. The 6 galaxies, the three left-most of which we do not include in the fits, 
that are to the left are all early-types. The slopes in eqs. (5) and (7) are 
similar, but the intercepts are different. This difference could be due to the 
assumptions made in deriving the two equations. In deriving equation (5) we 
assumed that mass of the dark matter halo is equal to the total gravitational mass of 
the galaxy. This assumption affects the intercept of the resulting relation. For example, 
if the mass of the dark matter halo is 90\%  of the total gravitiatonal mass, a 
reasonable assumption to make, then the intercept in eq. (5) would increase by 0.07, 
not enough to make the intercepts comparable.  On the other hand, the reason of 
the difference could be even more fundamental and has to do with the nature 
of the simulations from which eq. (6) is derived (di Matteo et al. 2003) 


\subsection{$M_{bh}-\sigma $ Relation:}
We plot in Fig. 2 $\log (M_{bh}/10^{8}M_{{\odot}})$ versus 
$\log (\sigma /200\ km\ s^{-1})$ for the 89 galaxies from Table 1.
The data can be fit by a straight line:
\begin{equation}
\log \frac{M_{bh}}{10^{8}\ M_{\odot}}=(0.22\pm 0.06)+(4.60\pm 0.31) \log \frac{\sigma}{200~km~s^{-1}}
\end{equation}
We estimate the intrinsic dispersion to be $\sim 0.50$ dex. The slope 
and normalization derived here agree, within the errors, with those of 
Gultenkin et al. (2009), Hu (2009), Graham (2008), Tremaine et al. (2002), 
and Merritt \& Ferrarese (2001), however, it is shallower than that 
reported by Beifiori et al. (2012) and McConnell \& Ma (2013). 
The relation is also in agreement with theoretical models 
(Di Matteo et al. 2003; Di Matteo et al. 2007). 

The amount of intrinsic scatter in our $M_{bh}-\sigma$ is less than that 
for the $M_{bh}-v_c$ above. This is in contrast to the values reported in 
Volonteri et al. (2011) in which the authors argue that the comparable 
intrinsic scatter in the $M_{bh}-\sigma$ and the $M_{bh}-v_c$ relations 
indicates that there is a trend of some sort between $M_{bh}$ and $v_c$. 
Our sample, consisting of 89 galaxies,  is larger than the one used in 
Gultekin et al. (2009a), 49 galaxies, and Volonteri et al. (2011), 25 galaxies 
used by Kormendy et al. (2011). In contrast, to Fig. 1, galaxies in Fig. 2 of 
all morphological types are spread out over roughly the same region. 


\subsection{Indirect $M_{bh}-v_c$?}

The large scatter in the $M_{bh}-v_c$ plot (Fig. 1) is contrary to what was found 
in some earlier studies (Ferrarese 2002; Baes et al. 2003). Baes et al. (2003) have 
calculated the blackhole masses indirectly by relying on the $M_{bh}-\sigma$ 
relation and then plotted $M_{bh}$ against $v_c$. They used the $M_{bh}-\sigma$ 
relation of Tremaine et al. (2002) with their $v_c-\sigma$ relation to derive a 
$M_{bh}-v_c$ relation. It is important to keep in mind that Baes et al. (2003) did  
{\it not} fit $M_{bh}$ and $v_c$. Our result here corroborates recent work by 
Beifiori et al. (2012) and Kormendy et al. (2011), but using a larger sample of 
dynamical blackhole masses.

To circumvent biases that could affect analysis relying on small samples, we use 
the very large sample of Ho (2007) to show that there will be a significant scatter 
in the $M_{bh}-v_c$ plots, even when using a $M_{bh}-\sigma$ relation, which 
by itself could introduce a relation which is not necessarily there (cf. our discussion
above of Ferrarese (2002) and Baes et al. (2003).  Ho (2007) presents $\sigma$ and 
$v_c$, together with other information, on 792 galaxies to study the 
$v_c-\sigma$ relation. Given the $\sigma$'s in Ho (2007), we use 
our $M_{bh}-\sigma$ relation, eq. (8), to calculate the 
corresponding blackhole masses for all the 792 galaxies. The errors on $\sigma$ 
and the uncertainties on the slope and normalization of the $M_{bh}-\sigma$ 
relation are propagated in quadrature together with the intrinsic scatter in 
$M_{bh}$ at constant $\sigma$ to calculate the uncertainties on the derived 
$M_{bh}$'s. We then plot the calculated masses against the corresponding $v_c$ 
for the 792 objects (Fig. 3).  The green circles are "kinematically normal" 
galaxies (Ho 2007) with $1 \lesssim \frac{v_c}{\sigma} \lesssim 2$.  We also 
overplot eq. (5), which is based on the observed correlation between $M_{bh}$ 
and $M_{DM}$ (Bandara et al. 2009), and eq. (7), which is based on the theoretical 
correlation between $M_{bh}$ and $M_{DM}$ (di Matteo et al. 2003). As a visual 
aid, the horizontal line marks the blackhole mass obtained from eq. (8) 
for $\sigma=100$~km~s$^{-1}$. 

There is a significant amount of scatter with many galaxies occupying a 
wide swath with $100 \lesssim v_c \lesssim 400$~km~s$^{-1}$ and 
$2\times 10^6 \lesssim M_{bh} \lesssim 3\times 10^8$~M$_\odot$; 
roughly the same region bounded by eqs. (5) and (7). Most of the galaxies in this 
region are "kinematically normal" (Ho 2007).  The majority of galaxies in this region 
are spirals (Sa to Sc, Fig, 3d-e). Moreover, for any given morphological type, the 
spirals have the highest percentage of occupying this region. However, the region 
has quite a large scatter, roughly an order of magnitude, in $M_{bh}$ for a given 
$v_c$. This makes using it as a kind of a relation between the mass of central 
blackhole and the circular velocity of the host galaxy problematic, or at least not 
as beneficial as the $M_{bh}-\sigma$ relation, which is used as probe of the 
co-evolution, in terms of redshift and morphology, of galaxy 
spheroids and the supermassive blackholes. 

Apart from showing that masses of the supermassive blackholes are poorly 
correlated with the circular velocities of their host galaxies, Fig. 3 uncovers a few  
interesting points. There are virtually no galaxies that have high circular velocities 
but low blackhole masses (empty lower right corner), whereas low circular velocity 
apparently places no constraints on the mass of the blackhole (the vertical scatter with 
$30 \lesssim v_c \lesssim 150$~km~s$^{-1}$). The first part is expected: 
A high circular velocity is related to a more massive halo, and a more 
massive halo leads to a deeper potential well, and hence more infalling material 
that eventually accretes onto the central supermassive blackhole. 

We address the second part through studying the $M_{bh}-v_c$ plot, Fig. 3a, by 
morphological type (Fig. 3b  to 3e). Many elliptical and lenticular galaxies have 
high blackhole masses, $M_{bh} \gtrsim 10^7$ M$_\odot$, but with circular velocities 
as low as $25$ km~s$^{-1}$ and as high as $500$~km~s$^{-1}$. They form the 
horizontal scatter in Figs. 3b \& 3c. These galaxies have 
$100 \lesssim \sigma \lesssim 300$ km~s$^{-1}$.  The horizontal scatter 
shows that the blackhole mass and the circular velocity of the host galaxies are 
not correlated in any way. However, one must be careful in the interpretation. 
The circular velocities used here are based on integrated \ion{H}{1} velocity 
profiles, which are less than velocities obtained from optical rotation curves 
for early type galaxies (see Ho 2007). The discrepancy is almost a factor of 2. 
This can explain the spread of early type galaxies (morphological index 
$\lesssim 0$)  to the left in Figs. 3b \& 3c, and also in Fig. 3d (Sa \& Sb spirals). 
One the other hand, the points in the lower part of Fig. 3f are very late type, 
bulgeless galaxies whose stellar velocity dispersion is 
$\lesssim 50$~km~s$^{-1}$ and most of it is contributed by the rotation of 
the disk. 

It seems that the only galaxies that follow some sort of a correlation are the  
kinematically normal galaxies. However, the reasoning here seems circular.  This 
could be another manifestation of the "cosmic conspiracy" discussed in  
Kormendy et al. (2011). The kinematically normal galaxies by definition have 
$1 \lesssim \frac{v_c}{\sigma} \lesssim 2 $ with an average 
$\frac{v_c}{\sigma} \approx \sqrt{2}$. This fact, taken with the $M_{bh}-\sigma$ 
relation (eq. 8), means that $M_{bh} \propto v_c^4$. If we fit the 
616 kinematically normal galaxies in Fig. 3a, we get:
\begin{equation}
\log \frac{M_{bh}}{10^{8}M_{{\odot}}}=(-0.55\pm 0.02) + (4.07\pm 0.16) \log \frac{v_{c}}{200\ km\ s^{-1}}, 
\end{equation}
with no intrinsic dispersion, since it was already included in the calculation of the 
error on the derived $M_{bh}$. If this intrinsic scatter is not included, then the intrinsic 
dispersion in eq. (9) is $\sim 0.9$ dex. The slope is consistent with that expected
from a $M_{bh}-v_c$ relation via a $M_{bh}-\sigma$: the expected slope is $\sim 4$. 
The intrinsic scatter is significantly less than when fitting using dynamical $M_{bh}$ as 
in eq. (2).  The intrinsic scatter would disappear if we include the intrinsic dispersion of our 
$M_{bh}-\sigma$ relation in the errors of the calculated black hole masses. This is 
expected intrinsic scatter in this case is build in, and it behaves like another source of error 
since it dominates the measurement errors.

\section{Conclusions}
This paper highlights the importance of the need to use direct values of the quantities 
being used to study the relation. We attempted here to use dynamical measurements of 
blackhole masses since these measurements are "clean" in the sense that they do not 
depend on the properties of the host galaxy. Another issue that deserves similar 
attention is the getting similarily "clean" measurements of the circular velocities. 
Circular velocities depend on the where they are being measured in the galaxy, 
and the conversion between the circular velocity of the galaxy and the circular velocity 
of the halo is model dependent. In this study to took the pragmatic approach used by 
previous authors (e.g., {\bf Sun et al. 2013}; Beifiori et al. 2012; Kormendy et al. 2011) and implicitly assumed 
that all circular velocities are those of the dark matter halo. 

Using this assumption, we found that the correlation between the masses of the supermassive 
blackholes and the circular velocities is very weak. By {\it extrapolation}, we conclude that the 
correlation between the blackholes and the dark matter halo is also weak in the 
present-day Universe. Volonteri et al. (2011) argued the these two components of galaxies 
could have been more correlated in the past. Blackholes and dark matter halos are 
bound to be correlated somehow. The deep potential wells of massive dark matter 
halos are needed for the material to accumulated in the center and form a blackhole. 
The existence of bulgeless spiral galaxies with supermassive blackholes presents 
a challenge to the no $M_{bh}-v_{c}$ correlation. On the other hand, a $M_{bh}-v_c$ 
correlation presisting to the present would imply the existence of very massive blackholes 
in the cores of dark matter halos of galaxy clusters (Kormendy et al. 2011).

\acknowledgments{We thank the anonymous referee for helpful comments, and 
the scientific editor for pertinent suggestions. BMS wishes to thank the Abdus Salam International Center for 
Theoretical Physics (ICTP) for hospitality, and P. Monaco and P. Salucci for helpful 
discussions.}

\begin{figure}
\hspace{-0.7in}
\psfig{figure=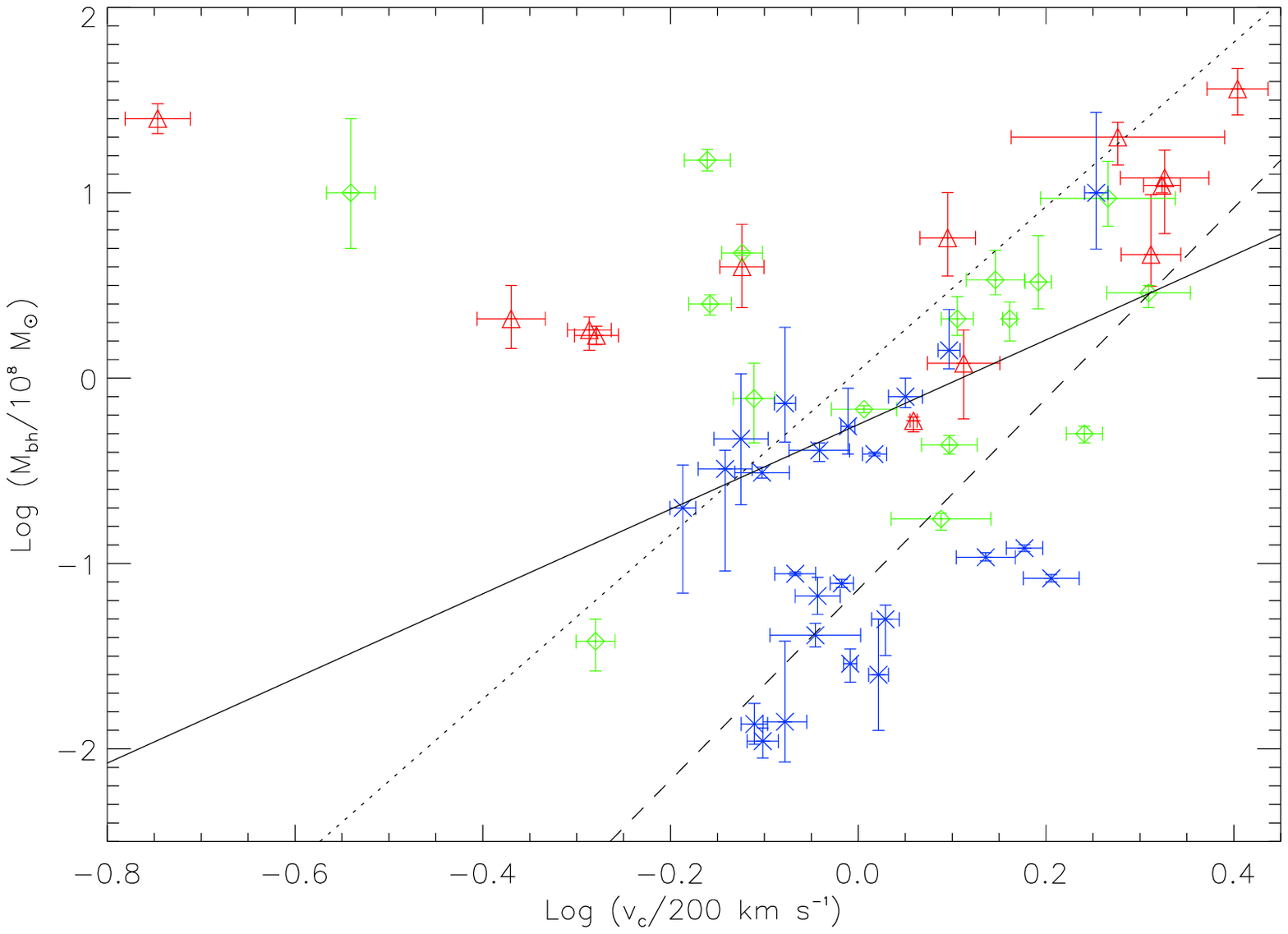}
\caption{$\log(M_{bh}/10^8 M_\odot)$ vs $\log(v_c/200~km~s^{-1})$ for 
the 53 galaxies in Table 1 that have dynamical blackhole mass and host 
galaxy circular velocity measurments. The ellipticals are the red 
triangles, the lenticulars are the green rhombii, and the spirals are the blue 
crosses. The solid line is the result of the fit, eq. (2).  The 3 galaxies with 
$v_c \lesssim 100$~km~s$^{-1}$ are not include in the fit. The dashed line is 
the semi-observational $M_{bh}-v_c$ relation eq. (5),  and the dotted line is 
eq. (7),  a theoretical $M_{bh}-v_c$ relation.} 
\end{figure}

\pagebreak

\begin{figure}
\hspace{-0.7in}
\psfig{figure=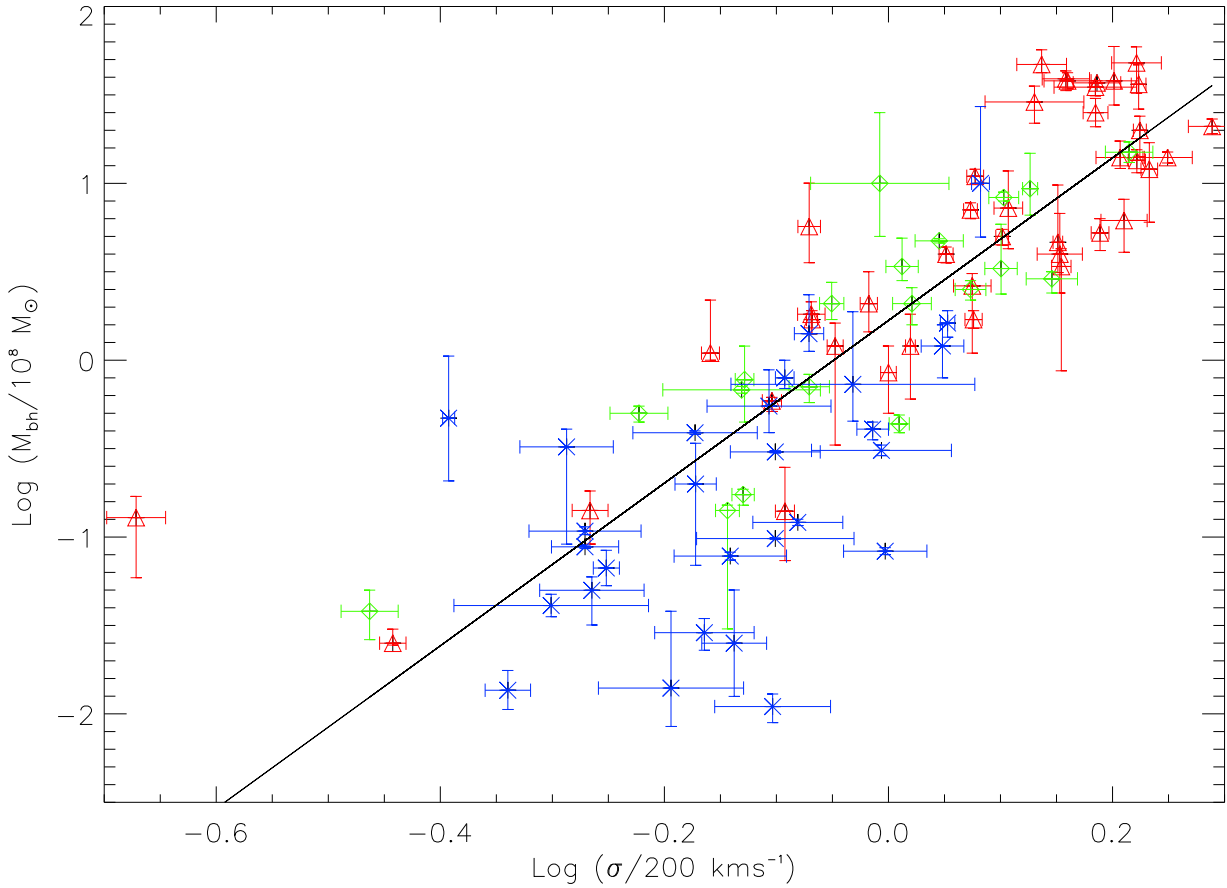}
\caption{$\log(M_{bh}/10^8 M_\odot)$ vs $\log(\sigma/200~km~s^{-1})$ for 
the 89 galaxies in Table 1 that have dynamical blackhole mass and host 
galaxy stellar velocity dispersion measurments. The ellipticals are the red 
triangles, the lenticulars are the green rhombii, and the spirals are the blue 
crosses. The solid line is the result of the fit, eq. (8).}
\end{figure}


\pagebreak
\begin{figure}
\vbox{
\hbox{
\psfig{figure=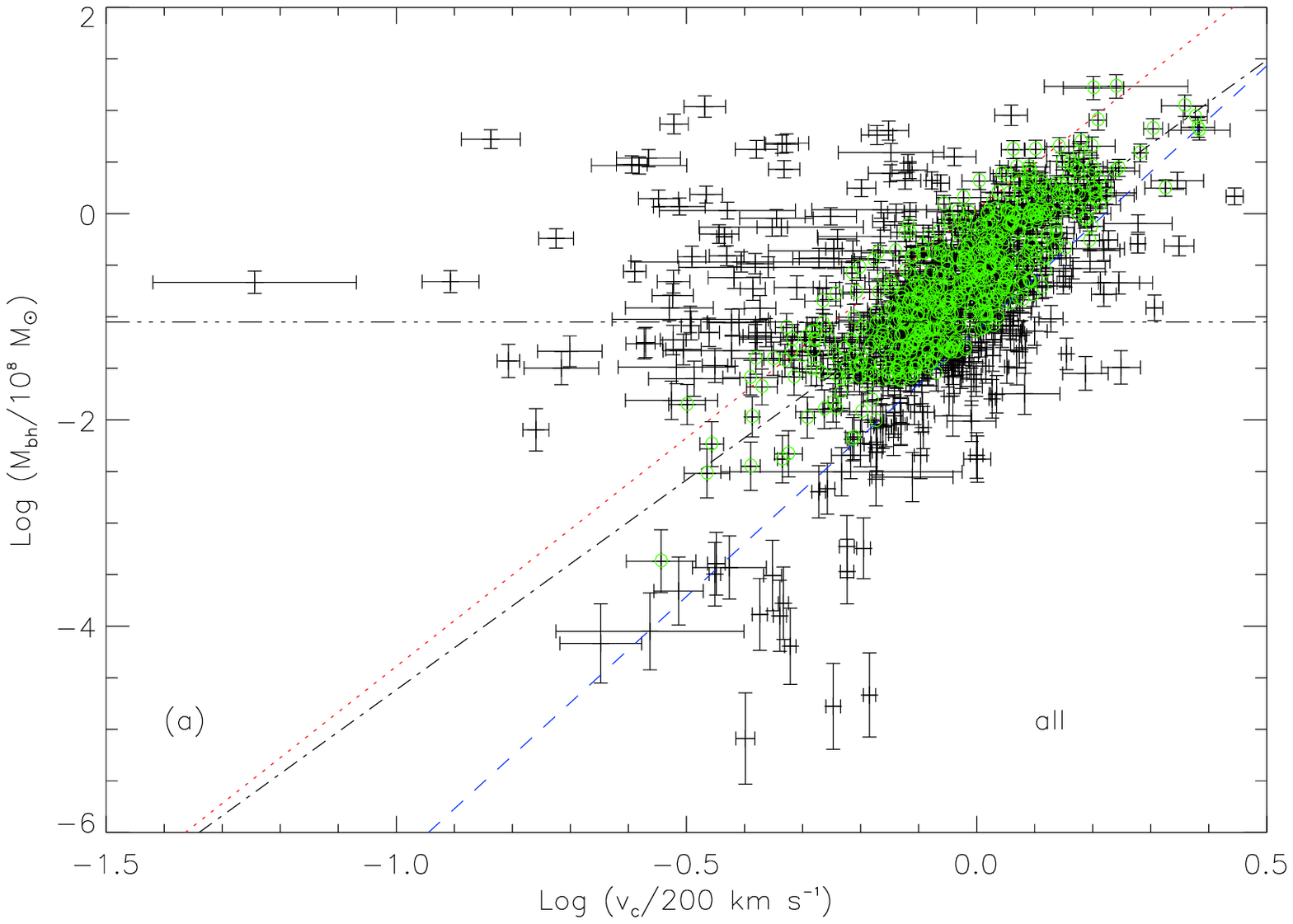,height=2in,width=2.5in}
\hspace{-0.3in}
\psfig{figure=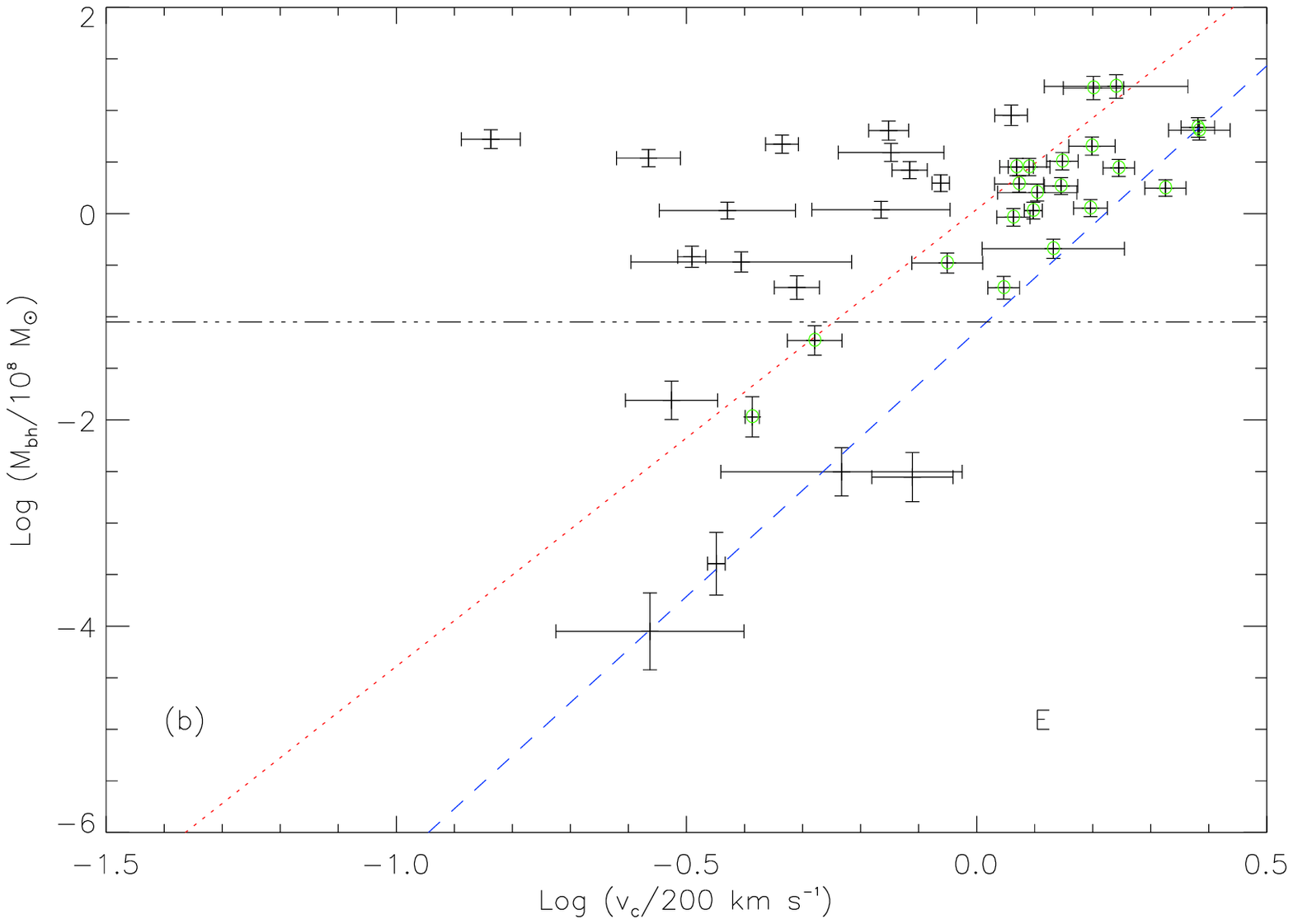,height=2in,width=2.5in}
\hspace{-0.3in}
\psfig{figure=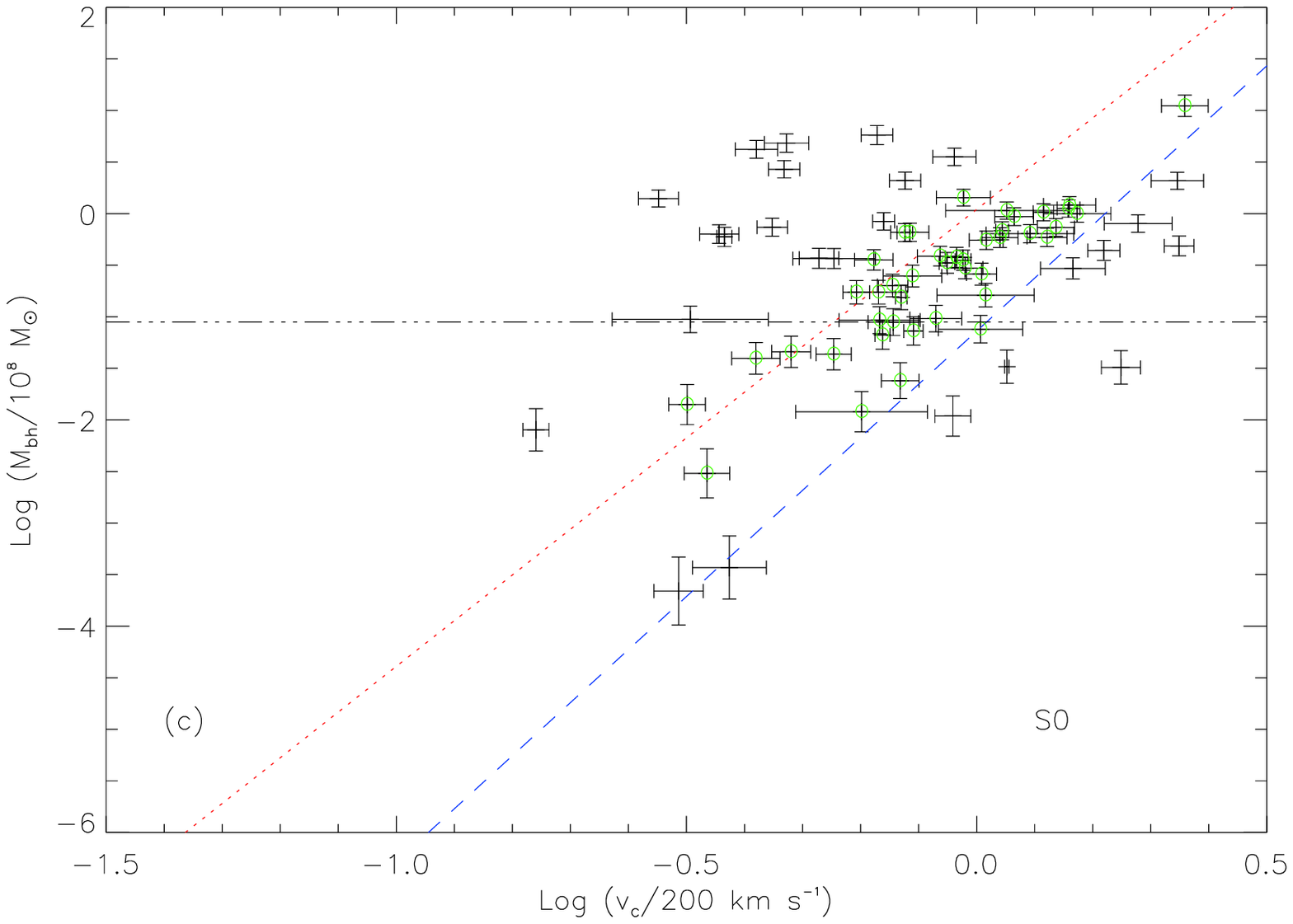,height=2in,width=2.5in}
}
\hbox{
\psfig{figure=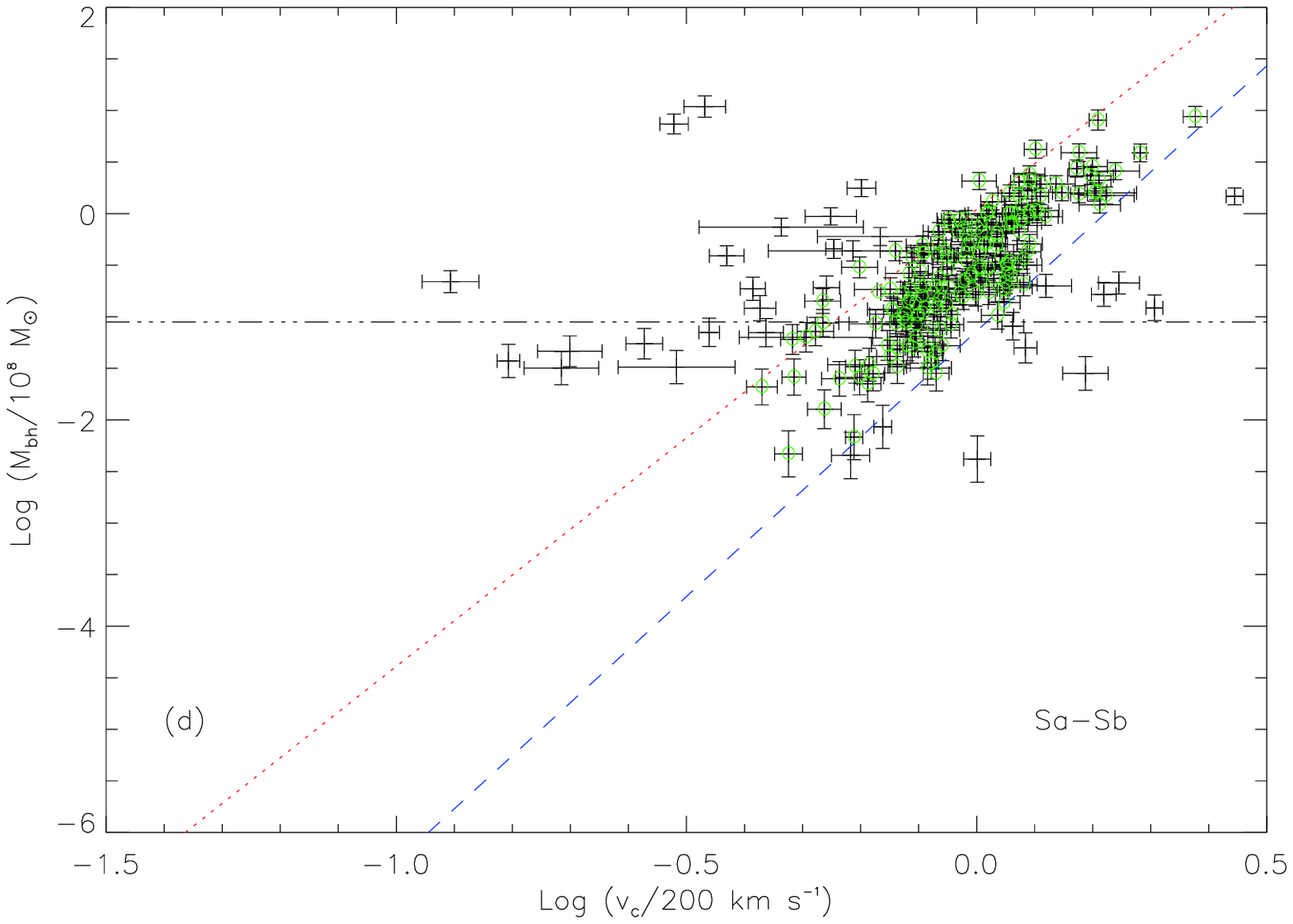,height=2in,width=2.5in}
\hspace{-0.3in}
\psfig{figure=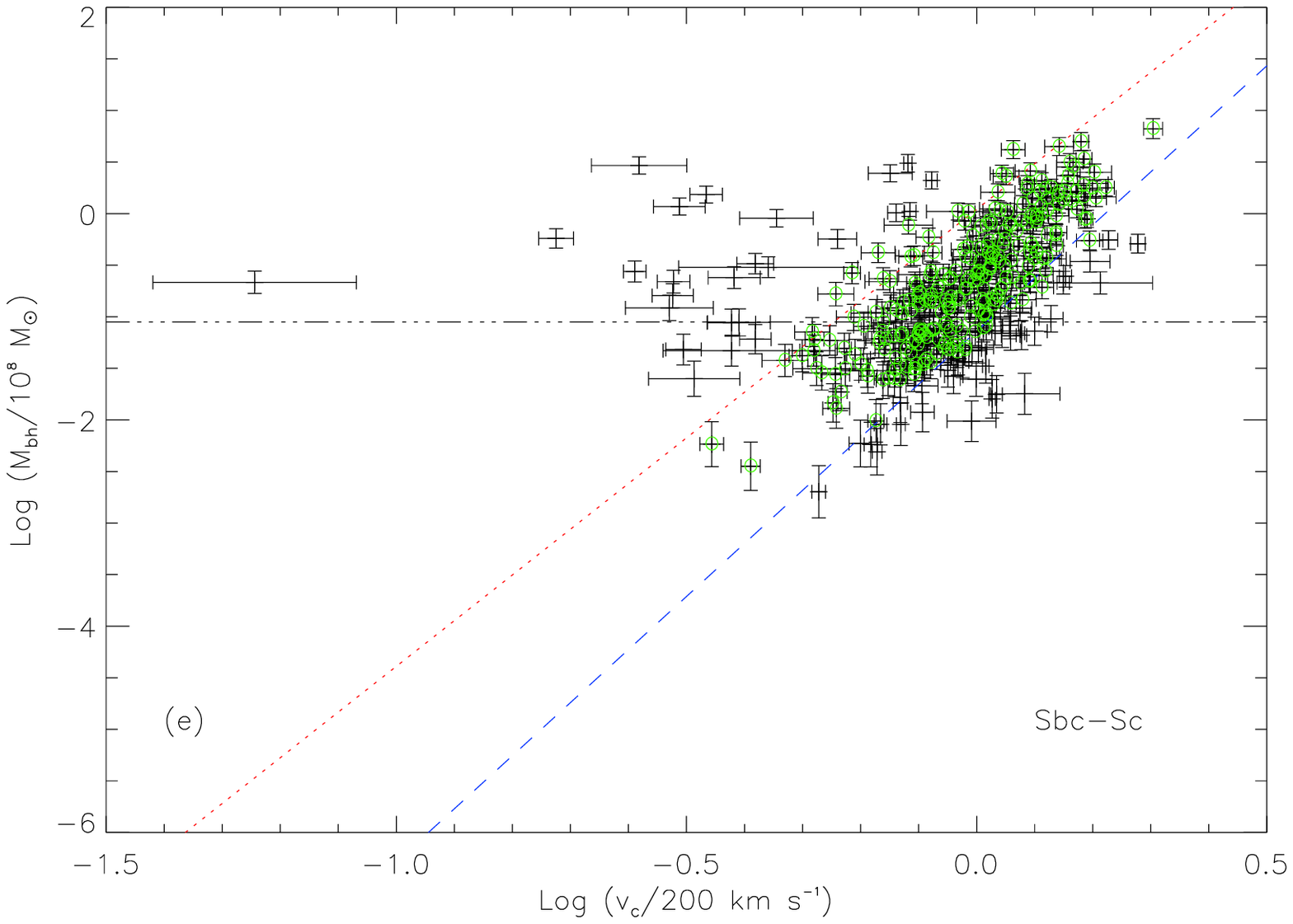,height=2in,width=2.5in}
\hspace{-0.3in}
\psfig{figure=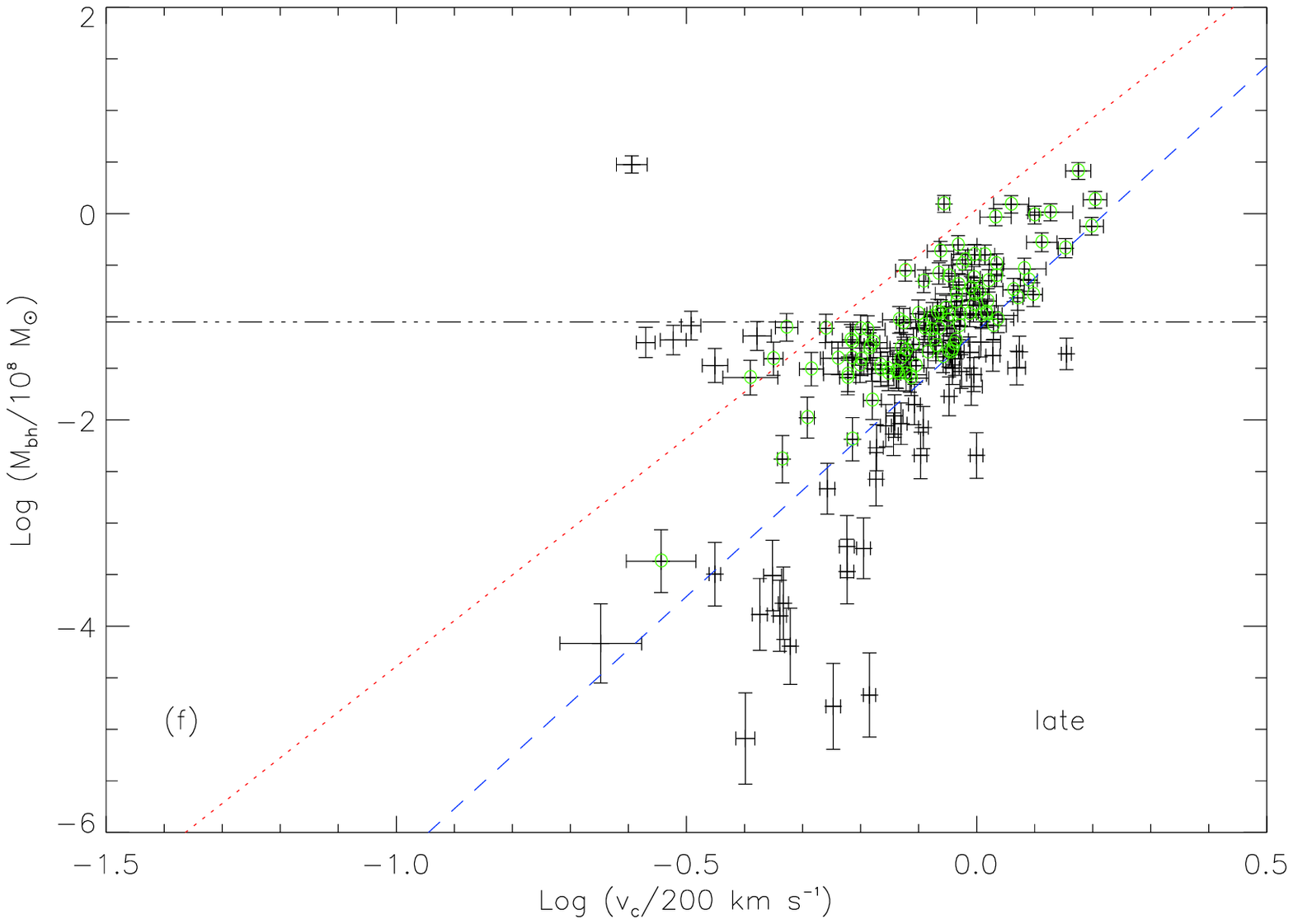,height=2in,width=2.5in}
}
\hbox{
\psfig{figure=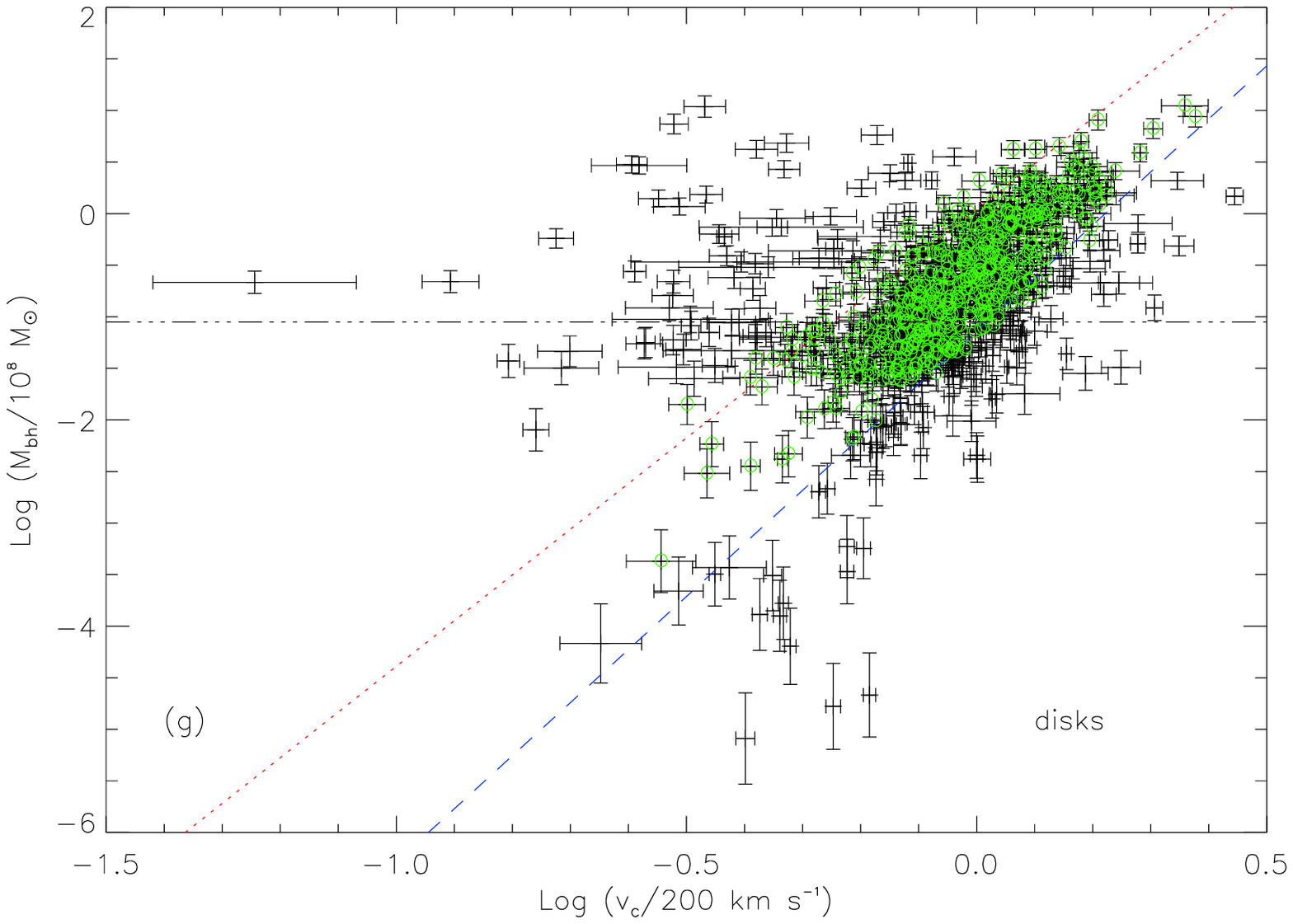,height=2in,width=2.5in}
\hspace{-0.3in}
\psfig{figure=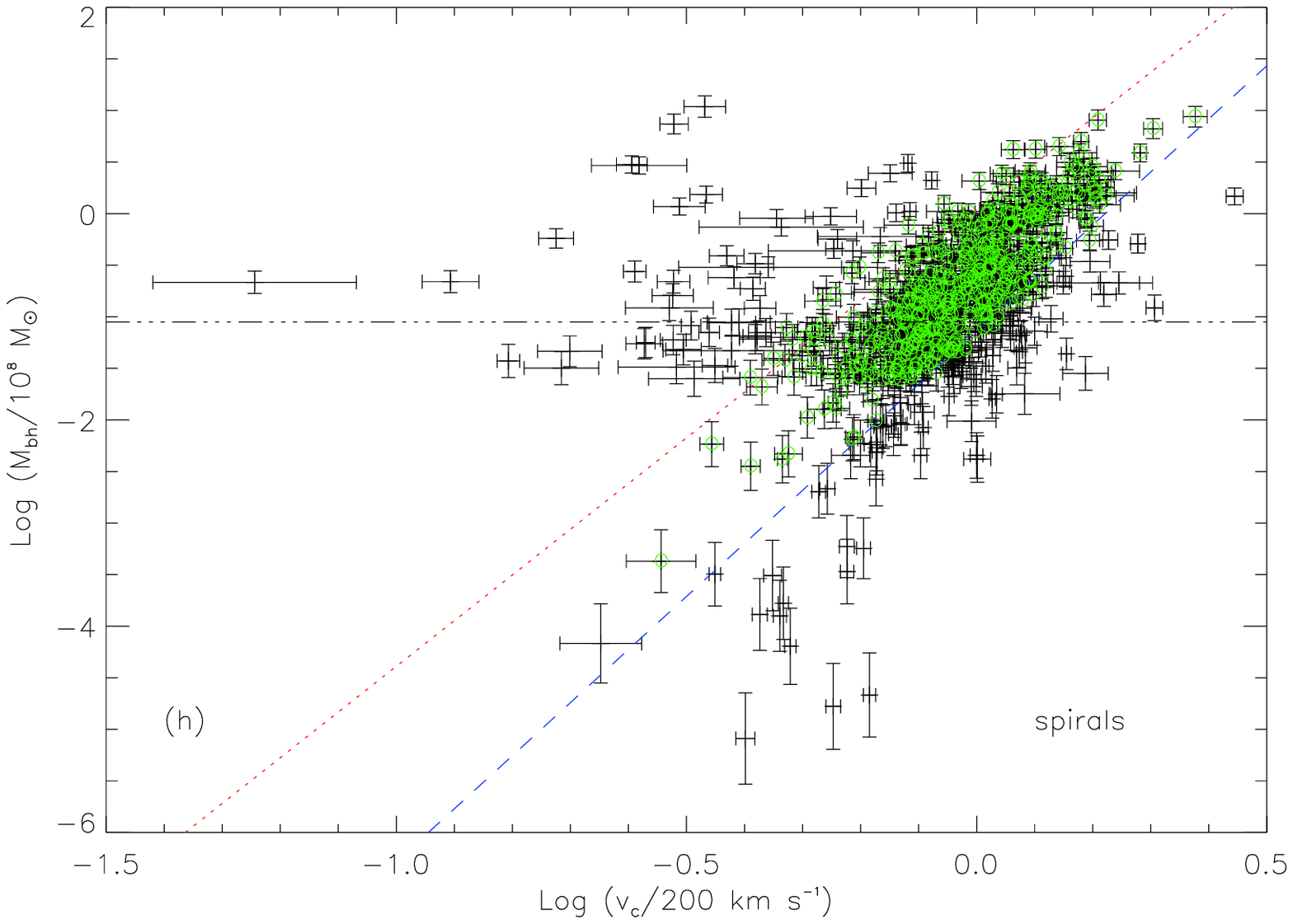,height=2in,width=2.5in}
}}
\caption{$\log(M_{bh}/10^8 M_\odot)$ vs $\log(v_c/200~km~s^{-1})$ for 
the 792 galaxies in the sample of Ho (2007). The blackhole masses have been calculated 
using from the stellar velocity dispersion according to eq. (8). (a): all types, (b): ellipticals, 
(c): S0 galaxies, (d): Sa \& Sb spirals, (e): Sbc \& Sc spirals, (f): later than Sc spirals, 
(g): all disk galaxies, and (h): all spiral galaxies. The blue dashed line is the 
semi-theoretical $M_{bh}-v_c$ relation eq. (5),  and the red dotted line is eq. (7),  
a theoretical $M_{bh}-v_c$ relation. Kinematically normal galaxies (see text for definition) 
are the green circles. The black dotted-dashed line in (a) is the fit to all 616 kinematically 
normal galaxies. The black horizontal line is the blackhole mass calculated from eq. (8) for 
$\sigma=100$ kms~s$^{-1}$.}
\end{figure}

\begin{table}
\caption{Galaxy Sample}
\begin{tabular}{cccccccc}
\tableline
\tableline
Galaxy	&	MORPH	&	D (Mpc)	&	A	&	$	M_{bh}/(10^6~M_\odot)					$	&	$	v_c (km~s^{-1})			$	&	$	\sigma (km~s^{-1})			$	&	Rf($M_{bh}$)	\\
(1)	&	(2)	&	(3)	&	(4)	&		(5)						&		(6)				&		(7)				&	(8)	\\
\tableline
Circinus	&	SB	&	2.8	&	Y	&	$	1.10	^{+	0.18	}_{-	0.23	}$	&	$	158.22	\pm	6.04	$	&	$	157.63	\pm	18.77	$	&	1-m	\\
IC 1459	&	E3	&	29.2	&	Y	&	$	2511.89	^{+	462.71	}_{-	462.71	}$	&	$	35.87	\pm	2.86	$	&	$	306.1	\pm	7.8	$	&	2-s	\\
IC 2560	&	SBb,p	&	41.4	&	Y	&	$	2.88	^{+	0.53	}_{-	0.66	}$	&	$	196.08	\pm	3.12	$	&	$	137	\pm	14	$	&	3-m	\\
MW	&	SBbc	&	0.008	&	Y	&	$	4.10	^{+	0.6	}_{-	0.6	}$	&	$	180	\pm	20	$	&	$	100	\pm	20	$	&	4-p	\\
NGC 0224	&	Sb	&	0.76	&	Y	&	$	141.25	^{+	71.55	}_{-	32.52	}$	&	$	249.84	\pm	6.73	$	&	$	169.87	\pm	5.12	$	&	5-s	\\
NGC 1023	&	SB0	&	11.4	&	N	&	$	43.65	^{+	5.03	}_{-	5.03	}$	&	$	250	\pm	17	$	&	$	204.48	\pm	4.23	$	&	6-s	\\
NGC 1068	&	Sb	&	15.3	&	Y	&	$	8.32	^{+	0.38	}_{-	0.38	}$	&	$	321.01	\pm	21.98	$	&	$	198.68	\pm	17.02	$	&	7-m	\\
NGC 1194	&	S0	&	55.5	&	Y	&	$	68.00	^{+	3	}_{-	3	}$	&	$	202.84	\pm	16.27	$	&	$	147.9	\pm	23.95	$	&	47-m	\\
NGC 1300	&	SBbc	&	20.7	&	N	&	$	73.00	^{+	69	}_{-	35	}$	&	$	167.09	\pm	4.37	$	&	$	185.9	\pm	46.59	$	&	8-g	\\
NGC 1332	&	S0	&	22.7	&	N	&	$	1500.00	^{+	200	}_{-	200	}$	&	$	138.12	\pm	7.77	$	&	$	328	\pm	16	$	&	49-s	\\
NGC 1399	&	E	&	21.1	&	N	&	$	1202.26	^{+	415.25	}_{-	830.49	}$	&	$	424	\pm	46	$	&	$	341.88	\pm	5.84	$	&	9-s	\\
NGC 2273	&	SBa	&	26.8	&	Y	&	$	7.80	^{+	0.4	}_{-	0.4	}$	&	$	192.09	\pm	5.46	$	&	$	144.5	\pm	16.7	$	&	47-m	\\
NGC 2748	& 	Sc	&	24.9	&	N	&	$	47.00	^{+	38	}_{-	38.4	}$	&	$	150	\pm	10	$	&	$	81	\pm	1	$	&	8-g	\\
NGC 2787	&	SB0/a 	&	7.5	&	Y	&	$	40.74	^{+	3.75	}_{-	5.63	}$	&	$	181.77	\pm	13.5	$	&	$	193.61	\pm	6.29	$	&	10-g	\\
NGC 2960	&	Sa 	&	75.3	&	Y	&	$	12.10	^{+	0.5	}_{-	0.5	}$	&	$	300.62	\pm	13.41	$	&	$	166	\pm	15.35	$	&	47-m	\\
NGC 2974	&	E4 	&	21.5	&	Y	&	$	169.82	^{+	19.55	}_{-	19.55	}$	&	$	105.22	\pm	5.67	$	&	$	238.23	\pm	4.21	$	&	11-s	\\
NGC 3031	&	Sb 	&	3.9	&	Y	&	$	79.43	^{+	18.29	}_{-	10.97	}$	&	$	224.51	\pm	9.38	$	&	$	161.61	\pm	3.1	$	&	12-g	\\
NGC 3079	&	SB(s)c	&	19.1	&	Y	&	$	2.51	^{+	1.74	}_{-	1.74	}$	&	$	210.19	\pm	5.15	$	&	$	145.66	\pm	9.71	$	&	13-m	\\
NGC 3115	&	S0 	&	9.7	&	N	&	$	933.25	^{+	429.78	}_{-	322.33	}$	&	$	369	\pm	61	$	&	$	267.6	\pm	4.13	$	&	14-s	\\
NGC 3227	&	SBa	&	17.5	&	Y	&	$	19.95	^{+	10.57	}_{-	21.13	}$	&	$	130.05	\pm	4.1	$	&	$	134.56	\pm	5.71	$	&	15-g,s	\\
NGC 3245	&	S0	&	20.9	&	N	&	$	208.93	^{+	43.3	}_{-	57.73	}$	&	$	290	\pm	5	$	&	$	209.91	\pm	8.36	$	&	40-g	\\
NGC 3379	&	E1 	&	10.3	&	N	&	$	120.23	^{+	49.83	}_{-	83.05	}$	&	$	259	\pm	23	$	&	$	209.23	\pm	2.1	$	&	16-g,s	\\
NGC 3384	&	S0	&	11.6	&	N	&	$	17.38	^{+	1.2	}_{-	2.4	}$	&	$	245	\pm	30	$	&	$	148.35	\pm	3.4	$	&	17-s	\\
NGC 3393	&	Sba,p	&	51.8	&	Y	&	$	30.90	^{+	2.13	}_{-	2.13	}$	&	$	157.93	\pm	10.61	$	&	$	197.14	\pm	28.35	$	&	18-m	\\
NGC 3414	&	SB0	&	25.2	&	N	&	$	251.19	^{+	28.92	}_{-	34.7	}$	&	$	139.02	\pm	7.26	$	&	$	236.75	\pm	7.47	$	&	11-s	\\
NGC 3585	&	S0	&	21.2	&	N	&	$	338.84	^{+	124.83	}_{-	62.42	}$	&	$	280	\pm	20	$	&	$	205.67	\pm	6.83	$	&	30-s	\\
NGC 3608	&	E2	&	22.9	&	N	&	$	208.93	^{+	86.59	}_{-	76.97	}$	&	$	85.36	\pm	7.15	$	&	$	192.08	\pm	3.4	$	&	17-s	\\
NGC 3998	&	S0	&	18.3	&	Y	&	$	288.40	^{+	26.56	}_{-	53.13	}$	&	$	407.61	\pm	41.79	$	&	$	279.82	\pm	14.74	$	&	19-g	\\
NGC 4026	&	S0	&	15.6	&	N	&	$	208.93	^{+	57.73	}_{-	43.3	}$	&	$	255	\pm	10	$	&	$	177.98	\pm	4.45	$	&	30-s	\\
NGC 4151	&	SAB	&	13.9	&	Y	&	$	32.36	^{+	7.45	}_{-	40.98	}$	&	$	144.27	\pm	9.54	$	&	$	103.22	\pm	9.91	$	&	20-g,s	\\
NGC 4258	&	SABbc  	&	7.2	&	Y	&	$	38.90	^{+	0.9	}_{-	0.9	}$	&	$	208.09	\pm	6.17	$	&	$	134.4	\pm	17.18	$	&	21-m	\\
NGC 4303	&	SBbc	&	16.1	&	Y	&	$	5.00	^{+	0.87	}_{-	2.26	}$	&	$	213.78	\pm	7.25	$	&	$	108.72	\pm	11.66	$	&	22-g	\\
NGC 4342	&	S0	&	17	&	N	&	$	330.00	^{+	190	}_{-	110	}$	&	$	311	\pm	10	$	&	$	252.07	\pm	8.39	$	&	42-s	\\
NGC 4374	&	E	&	18.4	&	Y	&	$	464.00	^{+	346	}_{-	183	}$	&	$	410	\pm	30	$	&	$	283.29	\pm	2.81	$	&	23-g	\\
NGC 4388	&	SB	&	19.8	&	Y	&	$	8.80	^{+	0.2	}_{-	0.2	}$	&	$	171.34	\pm	8.57	$	&	$	107.2	\pm	7.4	$	&	47-m	\\
NGC 4486	&	E	&	17.2	&	Y	&	$	3630.78	^{+	919.62	}_{-	1170.43	}$	&	$	507	\pm	38	$	&	$	334.44	\pm	5.05	$	&	24-g	\\
NGC 4486B	&	E	&	15.3	&	N	&	$	570.00	^{+	322.00	}_{-	269.00	}$	&	$	249	\pm	17	$	&	$	169.93	\pm	3.94	$	&	25-s	\\
NGC 4526	&	S0	&	16.5	&	N	&	$	473.00	^{+	13	}_{-	13	}$	&	$	150.41	\pm	7.52	$	&	$	222	\pm	11	$	&	50-g	\\
NGC 4564	&	E3	&	15.9	&	N	&	$	58.88	^{+	2.71	}_{-	8.14	}$	&	$	229	\pm	2	$	&	$	157.4	\pm	3.1	$	&	17-s	\\
NGC 4594	&	Sa	&	9.8	&	Y	&	$	1000.00	^{+	1000.00	}_{-	700.00	}$	&	$	358.48	\pm	10.32	$	&	$	241.65	\pm	4.41	$	&	26-s	\\
NGC 4596	&	SB0	&	16.7	&	N	&	$	77.62	^{+	33.96	}_{-	42.9	}$	&	$	154.85	\pm	7.88	$	&	$	148.8	\pm	2.85	$	&	10-g	\\
NGC 4649	&	E1	&	17.3	&	N	&	$	1995.26	^{+	367.54	}_{-	689.14	}$	&	$	378	\pm	99	$	&	$	335.3	\pm	4.45	$	&	17-s	\\
NGC 4736	& 	Sab 	&	4.9	&	Y	&	$	6.68	^{+	1.54	}_{-	1.54	}$	&	$	181	\pm	10	$	&	$	112	\pm	3	$	&	52-s	\\
NGC 4826	&	Sab 	&	6.4	&	Y	&	$	1.36	^{+	0.35	}_{-	0.34	}$	&	$	155	\pm	5	$	&	$	91.47	\pm	4.27	$	&	52-s	\\
NGC 4945	&	SB	&	3.8	&	Y	&	$	1.40	^{+	1.4	}_{-	0.7	}$	&	$	167.05	\pm	9	$	&	$	127.92	\pm	19.09	$	&	27-m	\\
NGC 5128	&	S0	&	3.5	&	Y	&	$	50.12	^{+	4.62	}_{-	5.77	}$	&	$	348.26	\pm	15.44	$	&	$	119.77	\pm	7.13	$	&	28-g	\\
NGC 5252	&	S0	&	96.8	&	Y	&	$	1000.00	^{+	921.03	}_{-	690.78	}$	&	$	57.6	\pm	3.44	$	&	$	196.46	\pm	27.92	$	&	29-g	\\
NGC 5576 	&	E3	&	27.1	&	N	&	$	181.97	^{+	29.33	}_{-	46.09	}$	&	$	103.4	\pm	5.57	$	&	$	170.68	\pm	4.82	$	&	30-s	\\
NGC 5846	&	E0	&	24.9	&	N	&	$	1096.48	^{+	100.99	}_{-	100.99	}$	&	$	421.24	\pm	19	$	&	$	239.03	\pm	4.14	$	&	11-s	\\
\tableline

\end{tabular}

\end{table}

\begin{table}
{Table 1 (Cont'd): Galaxy Sample}\\
\begin{tabular}{cccccccc}
\tableline
\tableline
Galaxy	&	Morph	&	D (Mpc)	&	A	&	$	M_{bh}/(10^6~M_\odot)					$	&	$	v_c (km~s^{-1})			$	&	$	\sigma (km~s^{-1})			$	&	Rf($M_{bh}$)	\\
(1)	&	(2)	&	(3)	&	(4)	&		(5)						&		(6)				&		(7)				&	(8)	\\
\tableline
NGC 7052	&	E4	&	71.4	&	N	&	$	398.11	^{+	210.84	}_{-	201.67	}$	&	$	150.35	\pm	8.1	$	&	$	284.42	\pm	13.22	$	&	31-g	\\
NGC 7457	&	S0	&	13.2	&	N	&	$	3.80	^{+	1.05	}_{-	1.4	}$	&	$	105	\pm	5	$	&	$	68.85	\pm	4.04	$	&	17-s	\\
NGC 7582	&	SBab	&	22	&	Y	&	$	55.00	^{+	26	}_{-	19	}$	&	$	194.99	\pm	3.32	$	&	$	156.46	\pm	19.93	$	&	32-g	\\
UGC 3789	&	Sa 	&	48.4	&	N	&	$	10.80	^{+	0.6	}_{-	0.5	}$	&	$	273.4	\pm	19.78	$	&	$	107.2	\pm	12.35	$	&	47-m	\\
\tableline
			&		&&	& Galaxies with no $v_c$&\\
\tableline
A1836	&	E	&	157.5	&	N	&	$	3900.00	^{+	400	}_{-	600	}$	&	\nodata					&	$	288	\pm	14	$	&	34-g	\\
A3565	&	E	&	54.4	&	N	&	$	1400.00	^{+	300	}_{-	200	}$	&	\nodata					&	$	322	\pm	16	$	&	34-g	\\
Cygnus A	&	E	&	240	&	Y	&	$	2884.03	^{+	597.67	}_{-	796.89	}$	&	\nodata					&	$	270	\pm	27.44	$	&	33-g	\\
IC 4296	&	cD	&	50.8	&	Y	&	$	1348.96	^{+	186.37	}_{-	217.43	}$	&	\nodata					&	$	333.24	\pm	5.87	$	&	34-s,g	\\
NGC 0221	&	E2	&	0.81	&	N	&	$	2.51	^{+	0.46	}_{-	0.06	}$	&	\nodata					&	$	72.21	\pm	1.96	$	&	35-s	\\
NGC 0524	&	S0,c	&	23.3	&	N	&	$	831.76	^{+	57.46	}_{-	38.3	}$	&	\nodata					&	$	253.46	\pm	7.78	$	&	36-s	\\
NGC 0821	&	E4 	&	24.1	&	N	&	$	85.11	^{+	29.4	}_{-	45.08	}$	&	\nodata					&	$	200.01	\pm	3.21	$	&	37-s	\\
NGC 1277	&	S0	&	73	&	N	&	$	17000.00	^{+	3000	}_{-	3000	}$	&	\nodata					&	$	333	\pm	17	$	&	48-s	\\
NGC 1316	&	SAB	&	20	&	N	&	$	162.18	^{+	26.14	}_{-	29.87	}$	&	\nodata					&	$	225.85	\pm	3.36	$	&	38-s	\\
NGC 1407	&	E	&	29	&	N	&	$	4700.00	^{+	900	}_{-	500	}$	&	\nodata					&	$	274	\pm	14	$	&	49-s	\\
NGC 1550	&	E	&	53	&	N	&	$	3800.00	^{+	400	}_{-	400	}$	&	\nodata					&	$	289	\pm	14	$	&	49-s	\\
NGC 2549	&	S0	&	12.3	&	N	&	$	14.13	^{+	0.98	}_{-	21.79	}$	&	\nodata					&	$	143.66	\pm	3.52	$	&	36-s	\\
NGC 2778	&	E2	&	22.9	&	N	&	$	14.00	^{+	8	}_{-	9	}$	&	\nodata					&	$	161.67	\pm	3.16	$	&	17-s	\\
NGC 3091	&	E	&	52.7	&	N	&	$	3700.00	^{+	100	}_{-	200	}$	&	\nodata					&	$	307	\pm	15	$	&	49-s	\\
NGC 3377	&	E5	&	11.2	&	N	&	$	109.65	^{+	75.74	}_{-	10.1	}$	&	\nodata					&	$	138.72	\pm	2.57	$	&	39-s	\\
NGC 3607	&	SA	&	19.9	&	Y	&	$	120.23	^{+	33.22	}_{-	49.83	}$	&	\nodata					&	$	223.5	\pm	9.8	$	&	30-s	\\
NGC 4261	&	E2	&	31.6	&	Y	&	$	524.81	^{+	96.67	}_{-	120.84	}$	&	\nodata					&	$	308.95	\pm	5.63	$	&	41-g	\\
NGC 4291	&	E2	&	26.2	&	N	&	$	338.84	^{+	78.02	}_{-	460.33	}$	&	\nodata					&	$	285.3	\pm	5.69	$	&	17-s	\\
NGC 4459	&	S0	&	16.1	&	N	&	$	70.79	^{+	11.41	}_{-	14.67	}$	&	\nodata					&	$	169.95	\pm	7.06	$	&	10-g	\\
NGC 4473	&	E5	&	15.3	&	N	&	$	120.23	^{+	35.99	}_{-	155.03	}$	&	\nodata					&	$	179.25	\pm	2.96	$	&	17-s	\\
NGC 4486A	&	E2	&	18.3	&	N	&	$	12.88	^{+	3.56	}_{-	10.09	}$	&	\nodata					&	$	42.61	\pm	2.58	$	&	43-s	\\
NGC 4552	&	E	&	15.9	&	Y	&	$	501.19	^{+	46.16	}_{-	57.7	}$	&	\nodata					&	$	252.65	\pm	3.28	$	&	11-s	\\
NGC 4621	&	E5	&	18.3	&	N	&	$	398.11	^{+	36.67	}_{-	45.83	}$	&	\nodata					&	$	225.15	\pm	3.23	$	&	11-s	\\
NGC 4697	&	E4	&	11.7	&	N	&	$	169.82	^{+	19.55	}_{-	31.28	}$	&	\nodata					&	$	170.92	\pm	1.95	$	&	17-s	\\
NGC 4742	&	E4	&	15.5	&	N	&	$	14.13	^{+	3.58	}_{-	6.18	}$	&	\nodata					&	$	108.32	\pm	3.99	$	&	44-s	\\
NGC 4751  	&	E/S0	&	27.7	&	N	&	$	1400.00	^{+	100	}_{-	100	}$	&	\nodata					&	$	355	\pm	18	$	&	49-s	\\
NGC 5077	&	E3,c	&	40.2	&	Y	&	$	724.44	^{+	350.3	}_{-	383.66	}$	&	\nodata					&	$	255.87	\pm	7.54	$	&	45-g	\\
NGC 5328	&	E	&	65.9	&	N	&	$	4800.00	^{+	1000	}_{-	1900	}$	&	\nodata					&	$	333	\pm	17	$	&	49-s	\\
NGC 5516	&	E	&	60.1	&	N	&	$	3500.00	^{+	200	}_{-	400	}$	&	\nodata					&	$	306	\pm	26	$	&	49-s	\\
NGC 5813	&	E1	&	32.2	&	N	&	$	707.95	^{+	65.2	}_{-	81.51	}$	&	\nodata					&	$	236.74	\pm	3.35	$	&	11-s	\\
NGC 5845	&	E3	&	25.9	&	N	&	$	263.03	^{+	42.39	}_{-	230.14	}$	&	\nodata					&	$	237.61	\pm	9.17	$	&	17-s	\\
NGC 6086	&	E	&	139.1	&	N	&	$	3800.00	^{+	1700	}_{-	1200	}$	&	\nodata					&	$	318	\pm	16	$	&	51-s	\\
NGC 6251	&	E2	&	107	&	Y	&	$	616.60	^{+	170.37	}_{-	255.56	}$	&	\nodata					&	$	324.57	\pm	15.4	$	&	46-g	\\
NGC 6264	&	S	&	145.4	&	Y	&	$	30.30	^{+	0.5	}_{-	0.4	}$	&	\nodata					&	$	158.5	\pm	14.65	$	&	47-m	\\
NGC 6323	&	Sab 	&	110.5	&	N	&	$	9.80	^{+	0.1	}_{-	0.1	}$	&	\nodata					&	$	158.5	\pm	25.65	$	&	47-m	\\
NGC 6861	&	E/S0	&	28.9	&	N	&	$	2100.00	^{+	200	}_{-	200	}$	&	\nodata					&	$	389	\pm	19	$	&	49-s	\\
\tableline
\end{tabular}
{NOTES. Col. (2): Morphological classification (NASA/IPAC Extragalactic Database. Col. (3): Distances (see 
col. (5) for references). Col. (4): Active Galaxy. Col. (5): Blackhole masses from Hu (2009) McConnell \& Ma (2013) 
except for NGC 1300, NGC 2778; NGC 4342; NGC 4374;  and NGC 4945 (Graham 2008), NGC 4303 (Pastorini 2007),  
NGC 4486 Kormendy et al. (1996), and  NGC 4594 (Kormendy et al. 1988).  (8). Col. (6): Galaxy circular 
velocity (see text for references). Col. (7): Central velocity dispersion from HyperLeda. Col. (8): 
Original references for blackholes masses. Methods for derivation of blackhole masses in original 
references: g=gas dynamics; s=star dynamics; p=stellar proper motion; m=maser. Original references 
for blackhole masses: (1) Greenhill et al. (2003); (2) Cappellari et al. (2002); (3)  Tilak et al. (2008);
(4) Ghez et al. (2008);}\\ 
\end{table}
\begin{table}
(5) Bender et al. (2005); (6) Bower et al. (2001); (7) Lodato et al. (2003); 
(8) Atkinson et al. (2005); (9) Houghton et al. (2006); (10) Sarzi et al. (2001); (11) Cappellari et al. (2007);
(12) Devereux et al. (2003); (13) Yamauchi et al. (2004); (14) Kormendy et al. (1996); 
(15) Davis et al. (2006) and Hicks et al. (2008); (16) Gebhardt et al. (2001) and Shapiro et al. (2006); 
(17) Gebhardt et al. (2003); (18) Kondratko et al. (2008); (19) de Francesco et al. (2006); 
(20) Onken et al. (2007) and Hicks et al. (2008); (21) Hernstein et al. (1999); (22) Pastorini (2007);
(23) Maciejewski et al. (2001); (24) Harms et al. (1994) and Macchetto et al. (1997); 
(25) Kormendy et al. (1996); (26) Kormendy et al. (1988); (27) Greenhill et al. (1997);
(28) Neumayer et al. (2007); (29) Capetti et al (2005); (30) Gultekin et al. (2009b); 
(31) van der Marel (1998); (32) Wold et al. (2006); 
(33) Tadhunter et al. (2003); (34) Dalla Bonta et al. (2009);(35) Verolme et al. (2002);
(36) Krajnovic et al. (2009); (37) Richstone et al. (2004); (38)  Nowak et al. (2008);
(39) Copin et al. (2004); 
(40) Barth et al. (2001);
(41)  Ferrarese et el. (1998); 
(43) Nowak et al. (2007);
(42) Cretton et al. (1999) and Valluri et al. (2004); 
(44) Tremaine et al. (2002);
(45) de Fransesco et al (2008); 
(46) Ferrarese et al. (1999); 
(47) Kuo et al (2011);
(48) van den Bosch (2012);
(49) Rusli et al. (2011);
(50) Davies et al. (2013) and Gould (2013);
(51) McConnell et al. (2011).
(52) Kormendy \& Bender (2011).
\end{table}

\end{document}